# Science and Engineering Education: Who is the Customer?

Michael Courtney, PhD
U.S. Air Force Academy, 2354 Fairchild Drive, USAF Academy, CO 80840
Michael.Courtney@usafa.edu

Amy Courtney, PhD
BTG Research, PO Box 62541, Colorado Springs, CO 80962
Amy_Courtney@post.harvard.edu

*Abstract:*
As education systems move toward business models of operation, there is a strong tendency to misidentify the student as the customer. Misidentifying the student as the customer leads to interpretation of the course credit or degree as the product. The true product is the additional knowledge, skill, and ability that course credit and degree should represent. Consequences are potentially disastrous, because the notion that "the customer is always right" can lead to the perceived product (course credit or degree) meeting the desires of the misidentified "customer" (student) rather than the real product (value added to student) meeting the standards of the properly identified customers (future employers and taxpayers).

**I. Introduction**

A college education is now seen as a necessary part of fulfilling the American dream. More students are going to college, and the bulk of the financial burden falls on taxpayers through governmental support of public institutions and government supported financial aid to students at both public and private institutions. This support is often justified by the value and necessity of an educated workforce in the current and future economy [1].

Recent attempts to improve accountability and efficiency of educational institutions include a trend toward operating them like businesses rather than traditional education management styles [2,3]. Cost effectiveness is easily measured as the number of graduates per dollar spent. Financial resources are often distributed from government to institutions based on the number of enrolled students. Thus institutions face the temptation of viewing the student as the customer, because the student controls a much larger dollar amount than is personally invested [4]. In contrast, the real customers in education are those paying for it (taxpayers) and those depending on a quality product (future employers) [5,6].

Empathy of most teachers toward students also contributes to misidentifying the student as customer. Every teacher has experienced challenging college level courses where life circumstances or demanding instructors made a course difficult beyond their comfort level. In contrast, few teachers have had the experience of being an employer and suffering with poorly trained employees, bearing the expense of their incompetence until finally bearing the costs of termination and replacement.[1] Likewise, few teachers have experienced first hand the dismal failure of science and math education to prepare jurors to rationally evaluate the merits of a criminal case, especially if it involves nontrivial analysis of forensic evidence.[2] All teachers have been students. Few have been employers

---

[1] We've known a number of engineers who slipped through the quality control cracks of their educational programs and performed poorly. Most ended up performing tasks usually performed by non-degreed technicians until the employer recognized the situation and they were transferred or fired.

[2] A catastrophe is brewing here. When juries lack the competence to evaluate forensic testimony on the merits, they tend to pick an expert witness they like better, turning justice into a popularity contest.

or worked in technical fields outside of education.

## II. Examples
*A. Engineering*
One of us (MC) worked as a test engineer at a large internet hardware company, Cisco Systems. The job was to design an automated test system to ensure the wireless networking products met the company's standards, which in turn were designed to make sure our customers would be happy with product quality. This automated test system was deployed on the factory floor of subcontractors who manufactured the wireless networking products. Units that met the rigorous performance standards were shipped directly to customers, and units that failed were re-routed for repair or disposal. The subcontractor was paid based on the amount of product that met the standards and shipped.

Subcontractors often complained about the unfairness of the test system, especially early in a product cycle when they were still working out manufacturing kinks and product yield percentages were low. The company's insistence on only paying for quality product was costing the subcontractors money. Had the subcontractors determined their own quality standards, field failure rates would have been much higher, and customers would have been unhappy with product quality.

*B. Physics at West Point*
One of us (AC) taught the two-semester calculus-based physics sequence at the United States Military Academy at West Point. In return for a taxpayer investment of over $400,000 in educational expenses, West Point graduates spend at least five years as officers in the United States Army.

The physics sequence is required of all cadets, and is a pre-requisite for many later science and engineering courses. The Army requires significant scientific and problem solving skills of West Point graduates. These physics courses are expected to impart a number of specified learning objectives and contribute generally to the mathematical and scientific maturity of future Army officers.

Every student who fails one of these courses is reviewed for expulsion. Should the empathy of faculty with the plight of individual students struggling with a challenging subject outweigh the legitimate military and national security interests of the customer?

Recognizing the Army's needs, the department standardizes both the curriculum and objective assessment tools. Before teaching the introductory sequence, instructors go through an excellent new instructor training program stressing course objectives, pedagogy, and assessment of student learning. Each instructor is observed a number of times each semester through classroom visits by senior instructors and peers. Both the team leader and the course director keep a close eye on documented student accomplishment over the course of each semester. Documented accountability of the department toward instructors and the instructors toward students ensures that Army's requirements are met.

## III. Student Realities
The stated expectation of most educators is that college students should spend 2-3 hours in preparation each week for each class hour.[3] This equates to a 45-60 hour weekly work load for a 15 credit hour schedule. Many students are simply unwilling or unable to expend this level of time and effort and will take shortcuts on their academic work when they convince themselves that it won't hurt their grade.

The real benefits of higher education for a given student are roughly proportional to student effort averaged over time.[4] On average, if one student spends about half the effort of another for four years in college, the student making half the effort will unavoidably graduate with roughly half the added value. Does it make sense to let engineering students dictate how hard they should be working in school? Does it make sense for future chemists and forensic scientists to decide how hard they need to work in school?

---

[3]This is stated explicitly in many course descriptions and syllabi. However, given the rarity of instances where it is accurate, one wonders whether this comment is more for accrediting bodies than students.

[4]We would not be dogmatic about strict proportionality, but the results are certainly monotonic, and few students reach the point of diminishing returns before investing two hours of study time for each class hour.

Given the time pressures and myriad distractions, students will seldom work much harder than they think is needed to achieve their grade goals. Most students are one of two kinds:
- Students who start a course at full throttle, and then scale back efforts if early feedback suggests it would not compromise achieving grade goals.
- Students who start a course with mediocre effort and try later to increase efforts to the level perceived as necessary to achieve grade goals.

Few students maintain a level of effort far above what is necessary to achieve their grade goal in a course.

The quality of wireless networking products depends more on the standards of the automated test system than on the manufacturing process. Similarly, the abilities of graduates depend more on the level of academic rigor (how hard it is to reach a grade goal) than on the pedagogical prowess of the teacher.

**IV. Customer Interests**

Properly identifying the customer is key, because internal standards and assessment methods should focus on whether the product is meeting the expressed interests of the customer. Engineering and technology based businesses express customer interests in technical specifications and performance parameters of their products.

In education, taxpayer and employer interests are expressed by the careful development of curricula and standards designed to produce graduates who will be productive and capable members of society and the workforce. In the same way that high tech companies have an ethical obligation to ensure that products meet specifications, science and engineering educators must ensure that their students meet the standards described in syllabi and other documentation regarding course outcomes. In contrast, misapplication of the business model overemphasizes student happiness, retention, and ability to graduate in four years.

The customer of a wireless networking device doesn't care if it takes 20% more time to produce the product; they care only that the product meets expectations. The Army is concerned with the character, work ethic, and technical competence of West Point graduates. Likewise, employers don't care whether a graduate was happy in school or graduates in four years; they care about the abilities, maturity, and work ethic of the employee they hire.

Employers want a graduate who has lived up to challenging academic expectations so that they will live up to the tough expectations of the employer. Viewing the taxpayer and employer as customer stresses course outcomes and valid assessments of student accomplishment. Viewing the student as customer stresses student evaluations of teachers and retention (to keep the headcount high for the following semester). These views are at odds, because the student who sees himself as the customer will demand to be taught in a way pleasing to him rather than meeting future employer expectations.

In the same way that a computer network company needed to provide an automated test system to assure subcontractor quality, educational institutions should have assessment methods for independently verifying educational quality. If student evaluations and the number of students passing a course are primary indicators of instructor quality, it becomes too tempting for the teacher to lower standards.

**V. Consequences of Misidentifying Student as Customer**

The business rule that the customer is always right [7] and the common expectation that the customer has the authority in business relationships are at the heart of the negative consequences of misidentifying the student as customer [8]. Student evaluations of teachers are given the importance of customer satisfaction surveys and take on a disproportionate role in tenure, promotion, and reappointment processes.

The real customer in any situation actually has authority in their ability to dictate details to the producer, or take their business elsewhere. Should the student really have authority over the teacher in the classroom? The business model fails miserably here, because education
is a kind of apprenticeship or mentorship. To be successful, educators must set and apply the standards.

Viewing the student as customer elevates short-term student happiness over long-term improvement in abilities. Satisfaction comes later when the benefits of the difficult training are realized. How many successful college coaches would fare well in a survey of player satisfaction during training?

Viewing student as customer shifts teacher focus exclusively to a pedagogical role. The teacher is no longer an empowered gatekeeper with control over academic rigor and learning quality. This is analogous to removing the automated test system from the factory floor. This is the heart of grade inflation. This is why Johnny can't read. This is why more college graduates will lack the expected skills. Motivation is limited to the carrot; the stick is not available.

Motivating students requires both selling the beauty and benefits of knowledge and abilities in an area (the carrot), and awareness of potential failure with the attendant consequences (the stick). Misidentifying students as customers removes the motivation of the stick, because students shift the blame for failure to the instructor. In their minds, they paid for quality instruction. If they fail, the customer model inclines them to believe they are not getting what they paid for. Thus, under-performing students are deprived of the opportunity for proper introspection.

After a poor exam performance, we often encourage students to prepare more thoroughly by pointing to the obvious futility of repeating the same thing over and over and expecting a different result. However, for the customer student, the more likely application is to shop for a different instructor. The stick is an ineffective motivator for the student who thinks, "If you won't pass me, I'll find an instructor who will."

**VI. Discussion**
Taken to an absurd extreme, the business model of higher education can lead to quotas as legislatures pressure educational systems to produce the maximum numbers of graduates per dollar. If there is a de-facto quota for the number of students that pass a given course, the level of effort and rigor required to pass is in
the hands of the students. This ridiculous possibility can only be countered by establishing and maintaining standards based on the needs of the real customers (future employers and taxpayers).

Wireless networking product quality depends on rigidly maintaining established standards regardless of yield. Likewise, academic quality demands maintaining academic standards regardless of how many students pass or fail.

Students who view themselves as customers will have a difficult time identifying their employers as their customers in the real working world. Four years of internalizing the message that they can dictate standards to their teachers will skew their idea of who the boss really is in the working world. These students will be ill prepared for real, objective, externally imposed standards that exist in most of the professional world.

The time scale of real customer feedback is much longer in education than in technology businesses. Therefore, an educational business model needs effective ways to ensure product quality long before employers can express their dissatisfaction.

The path of least resistance is to pay attention to the squeaky wheels (students who pressure for lower standards,[5] and short-sighted government officials who want higher graduation rates [9]). Improvement requires empathy with the real customer. We like to see ourselves as future patients of students in health care programs, as business owners hiring an engineer, as parents of an aspiring math teacher's future students, or as a crime victim depending on the training of a forensic scientist to solve a serious crime. As taxpayers, we realize that the thoroughly trained and competent employees will bear a growing proportion of the tax burden if the number of properly educated employees continues to decline.

When faced with temptation to give passing grades to students who demonstrate substandard performance, empathy for the consequences for the real customer gives us strength to properly maintain standards. Proper application of the Golden Rule [10] demands that educators consider the needs of customers as well as needs of students. Taxpayers and future employers need competent students who have developed a good work ethic and have met rigorous academic standards. Students need competent teachers who assign grades according to documented accomplishment and objective standards. Students need teachers who care about their future success more than their short term happiness. Assigning grades that are not commensurate with demonstrated accomplishment violates the Golden Rule.

---

[5]Can I earn extra credit? Will this be on the test? It would only be fair to curve our grades. Can we have a multiple choice exam? I hate essay questions! You should drop the lowest exam grade. Other classes let us skip the final exam. Why not allow a formula sheet? That's not fair!

Undue weight on student course evaluations in promotion and tenure processes creates tension in faculty between upholding standards they believe are in the best interests of students and future employers and pleasing the students so as not to get punished on student course evaluations at the end of the semester. Carrell and West describe a situation at the United States Air Force Academy where all instructors have access to common exams prior to their administration and less experienced instructors tend to teach more directly to the exam and are rewarded with positive course evaluations; whereas, more experienced instructors emphasize deeper learning which is needed in downstream courses and tasks but tend to perform more poorly on student course evaluations.[11] Yunker and Yunker also find significant negative associations between student course evaluations in an introductory course and student performance in a downstream course.[12]

It is not sufficient to bring students as far as possible with the effort they are willing to expend. Students must meet a rigid and carefully considered standard to ensure success in subsequent courses and careers. Teachers have a moral and ethical duty as gatekeepers not to pass them until they do.

Educators need to make both personal and institutional efforts to better understand employer's needs in graduates. How many science and math teachers get regular feedback on the quality of students ultimately hired by engineering firms? How many take the time to listen to the horror stories of employers regarding inadequately trained graduates and the time and expense involved in rectification? Empathy for the customer is empathy for the future employer, and success in business demands proper understanding of customer needs.